\documentclass[reprint, amsmath, amssymb, aps, prl, superscriptaddress]{revtex4-2}%notitlepage
\usepackage{txfonts}
\usepackage{tabularx}
\usepackage{graphicx}
\usepackage{dcolumn}
\usepackage{bm, color}
\usepackage{amsmath}
\usepackage{amsfonts}
\usepackage{amssymb}
\usepackage{multirow}
\providecommand{\U}[1]{\protect\rule{.1in}{.1in}}

\newcommand{\vv}[1]{\boldsymbol #1}%????(??)???
\usepackage{hyperref}
\hypersetup{
 citecolor=blue,
 colorlinks = true,
 urlcolor = blue
}
\usepackage{physics}

\begin{document}

\title{
%Editor: We have provided a PDF that shows the tracked changes in your file as in a Word document. This method makes it easier for you to match the edited file with your original file and make any necessary edits to your file in your LaTeX program. Please let us know if you require further assistance.
Majorana
Spin Current Generation by
%Editor: Please ensure that the intended meaning has been maintained in the following edit here and elsewhere throughout the manuscript.
Dynamic
Strain}

\author{Yuki Yamazaki}
\affiliation{Department of Physics, Nagoya University, Nagoya 464-8602, Japan}
\author{Takumi Funato}
\affiliation{Center for Spintronics Research Network, Keio University, Yokohama 223-8522, Japan}
\affiliation{Kavli Institute for Theoretical Sciences, University of Chinese Academy of Sciences, Beijing, 100190, China}
\author{Ai Yamakage}
\affiliation{Department of Physics, Nagoya University, Nagoya 464-8602, Japan}

\date{\today}

\begin{abstract}

Majorana fermions that emerge on the surface of topological superconductors are charge neutral but can have higher-rank electric multipoles by allowing for account time-reversal and crystalline symmetries. Applying the general classification of these multipoles, we show that the spin current of Majorana fermions is driven by spatially nonuniform dynamic strains on the (001) surface of superconducting antiperovskite Sr$_3$SnO. We also find that the frequency dependence of the Majorana spin current reflects the energy dispersion of Majorana fermions. Our results suggest that the spin current can be a probe for Majorana fermions.

\end{abstract}

\maketitle
%%%%%%%%%%%%%%%%%%%%%%%%%%%%%%%%%%%%%%%%%%%

%\textit{Introduction.---}
Majorana fermions (MFs) are charge-neutral relativistic particles moving in 3D space.
In addition, two types of MFs emerge in topological superconductors (TSCs) as gapless Andreev bound states \cite{Hu1526, Kashiwaya1641, Hasan3045, Qi1057, Tanaka011013, Sato076501, Haim2019a}.
One type is a spatially localized zero-dimensional MF, which appears at the ends of nanowires \cite{Sato020401,Kitaev131,Lutchyn077001,Oreg177002,Cook201105,Alicea076501,Mourik336} or in the cores of the vortices of TSCs \cite{Volovik609,Read10267,Sarma220502,Fu096407}.
MFs have been the subject of considerable research because of their potential application to fault-tolerant topological quantum computation with non-Abelian statistics \cite{Nayak1083,Aasen031016}.

Here, one can ask the following question: what are the physical phenomena unique to spatially extended 1D or 2D MFs? A typical example is a half-integer thermal quantum Hall effect on the surface of a TSC \cite{Sumiyoshi023602, Nomura026802, Shimizu195139}.
``Half-Integer'' is a peculiarity originating from the fact that MFs are heat carriers. The half-integer thermal quantum Hall effect is unique to MFs that are spread in 2D space and
%Editor: Please ensure that the intended meaning has been maintained in the following edit.
 uses the thermal gradient
as the ``driving force''.
Recently, the optical response of Majorana chiral edge modes has also been discussed \cite{He237002,HeL241109}. The frequency dependence of real-part optical conductivity is proportional to $\omega^2$, where it can be distinguished from trivial superconductors or insulators and Dirac chiral edge modes. Then, the driving force of MFs is an electromagnetic wave.
In superconductors, we cannot use an electric field to survey superconducting properties since charge conservation is broken.
The responses to acoustic waves have been studied extensively through measurements of ultrasonic attenuation \cite{Tsuneto402, KadanoffA1170} to investigate superconducting symmetry or measurements of the temperature dependence of a superconducting gap \cite{Bruno2005}.
Therefore, we can expect to be able to use dynamic strains for the driving force of MFs on the surfaces of 3D TSCs.

\begin{figure}%\begin{figure*}
	\centering
\includegraphics[scale=0.25]{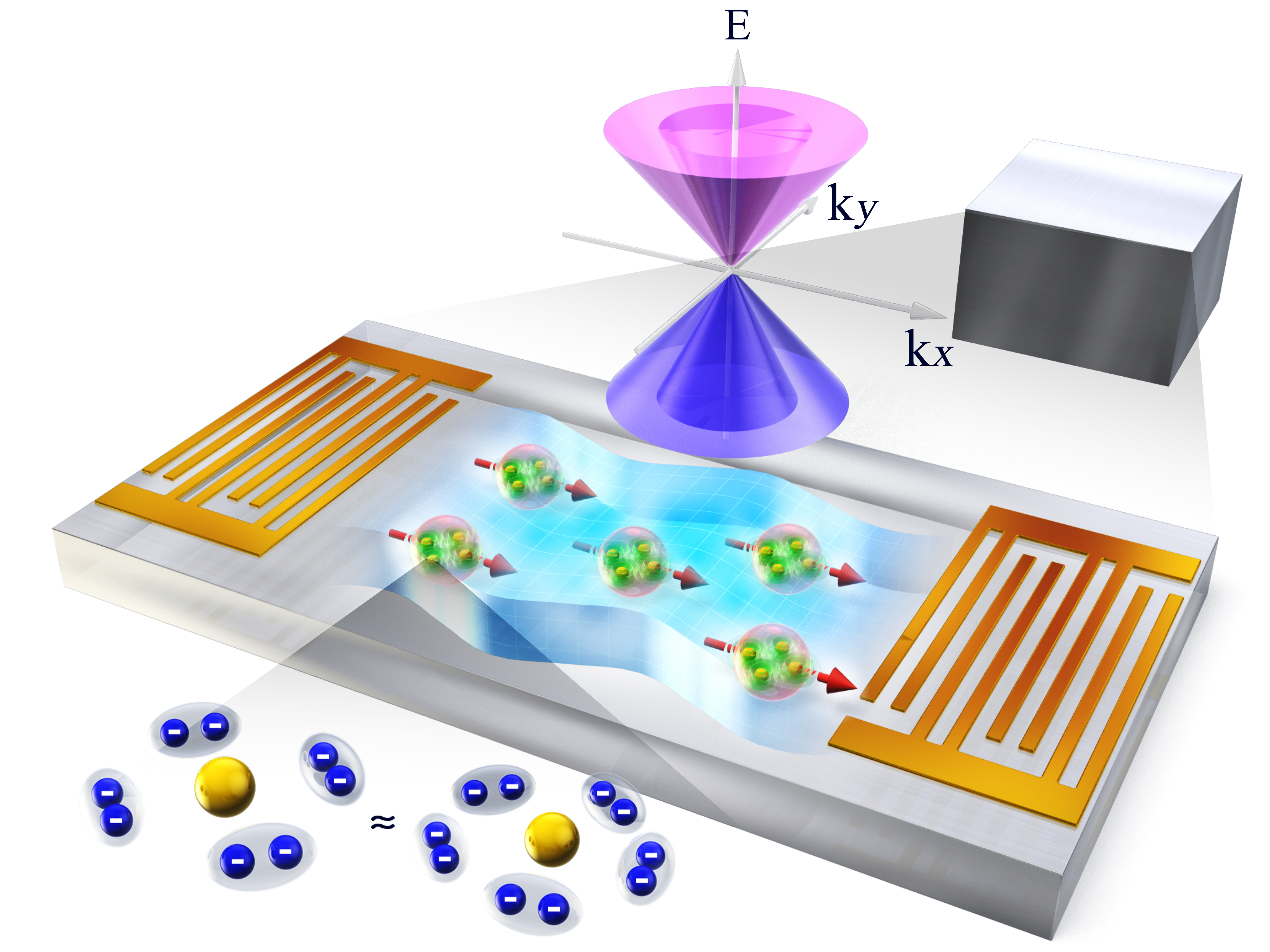}
\caption{This schematic shows that the Majorana spin currents flow on the surface of the 3D TCSC driven by a dynamic strain. The surface has four Majorana fermions that form double Majorana Kramers pairs (MKPs).
Double MKPs have electric quadrupoles coupled to a dynamic strain, and the spin current of Majorana electric quadrupoles is generated by a dynamic strain. %$j^{\alpha}_i$ denotes the Majorana spin current and $u_{ij}(\bm{x}) = \partial_i u_j(\bm{x})$ where $u_j(\bm{x}) \propto e^{i(\bm{q}\cdot \bm{x}-\omega t)}$ denotes the displacement field.
}
\label{3D}
\vspace{-5.5mm}
\end{figure}%\end{figure*}

\begin{table*}[t]
	\caption{Irreducible decomposition of the matrices, momentum, strain, and spin current in the $p4m$ model for the $A_{1u}$ superconducting pair potential in the bulk.
		The matrix is transformed by $g$ as $\eta \to D_g^\dag \eta D_g$.
		$(\eta_i,\eta_j)$ denotes $(\eta_i,\eta_j) \rightarrow (-\eta_j,\eta_i)$ by transformation of $C_{4z}$ and  $(\eta_i,\eta_j) \rightarrow (-\eta_i,\eta_j)$ by transformation of $\sigma(xz)$. Electric/magnetic and PHS denote time-reversal even/odd and particle-hole symmetry, respectively.
		$u_{ij}(\bm{x}) = \partial_i u_j(\bm{x})$ with $\vb*{u}$ the displacement field.
		$j^{\alpha}_{i}$ denotes spin $\sigma^\alpha/2$ current along the $i$th direction.
}

 \begin{tabular}{lllllllllllll}
      \\
  \hline\hline
  IR & \multicolumn{2}{l}{electric w/ PHS} & \multicolumn{2}{l}{electric w/o PHS} & \multicolumn{2}{l}{magnetic w/ PHS} & \multicolumn{2}{l}{magnetic w/o PHS} & momentum & strain & spin current
  \\
  \hline
  $A_{1}$ & & & $s_0\tau_0, s_0\tau_3$ & & $$ & & & & $k^2_x + k^2_y$ & $u_{xx} + u_{yy}$ \ \ \ \ & $j^{x}_{y}-j^{y}_{x}$
  \\
  $A_{2}$ & & & $$ & & $s_3\tau_0,s_3\tau_3$ & & & & $$ & $u_{xy} - u_{yx}$ & $j^{x}_{x}+j^{y}_{y}$
  \\
  $B_{1}$ & $s_3\tau_1$ & & & & & & $s_3\tau_2$ & & $k^2_x - k^2_y$ & $u_{xx} - u_{yy}$ & $j^{x}_{y}+j^{y}_{x}$
  \\
  $B_{2}$ & $s_0\tau_2$ & & & & & & $s_0\tau_1$ & & $k_x k_y$ & $u_{xy} + u_{yx}$ & $j^{x}_{x}-j^{y}_{y}$
  \\
  $E$ & $$ & & $(s_1\tau_1,s_2\tau_1)$ &  & $(\eta_5,\eta_6)$ & & $(\eta_1,\eta_2),(\eta_3,\eta_4)$ & & $(k_x,k_y)$ & $(u_{xz}, u_{yz})$ & $(j^{z}_{y},-j^{z}_{x})$
  \\
  \hline\hline
 \end{tabular}
\label{IR}
\end{table*}

When MFs have time-reversal symmetry, they do not appear alone but form Kramers pairs, which are called MKPs. Due to the time-reversal symmetry, a single MKP is stable against an electrical external field such as acoustic waves with lattice strains.
Previously, we derived the general effective theory for electromagnetic properties of ``double'' MKPs on a surface of a 3D TSC \cite{yamazaki07939,kobayashi224504,yamazaki073701}.
Double MKPs can have various electrical multipole degrees of freedom that are qualitatively different from those of ordinary fermions, and it has been shown that they can respond to static strains \cite{yamazaki073701}.
Double MKPs are always protected by crystalline symmetries in addition to time-reversal symmetry; thus, TSCs are specifically referred to as topological crystalline superconductors (TCSCs).

In this letter, we clarify the transport phenomena of double MKPs driven by spatially nonuniform dynamic strains on the surface of 3D TCSCs.
We consider the surface of the antiperovskite TCSC Sr$_3$SnO \cite{Oudah13617}, which has double MKPs with two electrical multipoles on its (001) surface \cite{Kawakami041026,yamazaki073701}.
We start with a $4 \times 4$ Dirac Hamiltonian with respect to the (001) surface point group (PG) symmetry.
%we construct the surface effective model for double MKPs that reflects the superconducting and crystalline symmetries of the 3D bulk TSC.
Then, we reveal the spin current of double MKPs, referred to as the ``Majorana spin current'', generated by spatially nonuniform dynamic strains on the surface (see Fig. \ref{3D}).
%In order to neglect the contribution of impurities, we consider a situation where a magnetization is presence in the perpendicular direction and a energy gap is opened in the dispersion of MKPs.
When there is no gap in the dispersion of double MKPs, we find that the Majorana spin current is caused by an intrinsic effect and does not depend on relaxation times. Alternatively, for the gapped case induced by surface magnetization, the Majorana spin current has two frequency regions that reflect the energy dispersion of double MKPs.
Our study suggests that the measurement of the Majorana spin current can be a new probe for Majorana fermion observation.

%Majorana spin current is a new phenomenon unique to MFs spread in two-dimensional space and caused by an external field beyond the thermal gradient without bulk excitation. Therefore, our study can be not only one of the probes for MFs observation but also the opportunity to accelerate the study of transport phenomena in MFs.And hence, Majorana spin current is clearly distinct from bulk or an accidental surface spin current.

%\textit{Electromagnetic properties of MKPs.---}
Before we consider the main topic, we discuss the electromagnetic properties of MKPs on the surface of a TSC. When there is only one MF on the surface, the single MF is strongly protected by particle-hole symmetry (charge-neutrality constraint), and it is extremely stable against any external fields. If a TSC has time-reversal symmetry, then the two MFs form an MKP. The single MKP is protected by time-reversal symmetry, and hence, it generally responds only to magnetic external fields and remains stable to electrical external fields. On the other hand, when crystalline symmetries are taken into account, there can exist double MKPs, and they can respond to an electrical perturbation that breaks the crystalline symmetry. The coupling of $N$ MKPs to a spatially uniform external field can be represented by Ref. \cite{yamazaki073701}
\begin{align}
 \hat{H}_{\mathrm{surf}, \mathrm{ex}} = - \hat{O} F,
 \
 \hat{O} =
 \frac{1}{2} \int d^2x \sum_{ss'} \hat{\psi}_{s}(\boldsymbol{x}) (A_F)_{ss'} \hat{\psi}_{s'}(\boldsymbol{x}),
 \label{OF}
\end{align}
where $\hat{\psi}_{s}(\boldsymbol{x}) \ (s=1,...,2N)$ are Majorana field operators and satisfy $\hat{\psi}^{\dagger}_{s}(\boldsymbol{x}) =\hat{\psi}_{s}(\boldsymbol{x})$. $A_F$ is conjugate to $F$ and should be an antisymmetric Hermite matrix since Majorana field operators obey $\{ \hat{\psi}_{s}(\boldsymbol{x}), \hat{\psi}_{s'}(\boldsymbol{x}') \} = \delta_{ss'}\delta^2(\boldsymbol{x} - \boldsymbol{x}')$. When $A_F$ is the $2 \cross 2$ matrix, with only one MKP, there exists only $A_F = \sigma_y$, which satisfies $\Theta_{\text{surf}} A_F \Theta^{-1}_{\text{surf}} = -A_F$ for the time-reversal operator $\Theta_{\text{surf}} = (-i \sigma_y K)$. $K$ is a complex conjugation, and hence, the MKP couples only to magnetic external fields. On the other hand, when $A_F$ is the $4 \cross 4$ matrix, with double MKPs, $A_F$, which satisfies $\Theta_{\text{surf}} A_F \Theta^{-1}_{\text{surf}} = A_F$, can be formed, and the double MKPs can also be coupled to the electrical external field by Eq. (\ref{OF}).
Double MKPs have various electric multipoles depending on each crystalline symmetry of the surface, i.e., wall-paper groups (WGs), and superconducting symmetries. In a previous study \cite{yamazaki073701},
%Editor: Please ensure that the intended meaning has been maintained in the following edit.
we revealed the coupling between the electric multipoles and the spatially uniform static strain
for each WGs.

%\textit{Model.---}
We consider the antiperovskite superconductor Sr$_3$SnO as a concrete model. Antiperovskite $A_3BX$, $A=\mathrm{Ca}, \mathrm{Sr}, \mathrm{La}$, $B=\mathrm{Pb}, \mathrm{Sn}$, $X=\mathrm{C}, \mathrm{N}, \mathrm{O}$ with the space group symmetry $Pm\bar{3}m$ (No.221) \cite{Widera1805,Nuss300}, has multiple
$J =3/2$ bands and is a candidate material for topological crystalline insulators due to the band inversion of two orbitals \cite{kariyado11, kariyado12, hsieh14}. In particular, Sr$_3$SnO has the potential to become a TSC with hole doping at temperatures below 5 K \cite{Oudah13617}.
Interestingly, it has been suggested that double MKPs can appear in the $(001)$ surface with $p4m$ WG symmetry when the superconductor symmetry is in the $A_{1u}$ representation in bulk with $O_h$ PG symmetry \cite{Kawakami041026}. In the following, we consider the double MKPs on the $(001)$ surface in the $A_{1u}$ state of Sr$_3$SnO and derive the Majorana spin current generated by spatially nonuniform dynamic strains.

To analyze the transport properties of double MKPs, we utilize the low-energy Dirac model for the (001) surface of Sr$_3$SnO with $p4m$ WG symmetry, which equals $C_{4v}$ PG symmetry.
The surface symmetry operations are given by $D_{\{C_{4z}| \vv{0} \}}= \frac{-1}{\sqrt{2}}(s_0\tau_3 - is_3\tau_3)$ and $D_{\{ \sigma(xz)|\vv{0} \}} = \frac{i}{\sqrt{2}}(s_2\tau_3 - s_1\tau_3)$ with $D_{g}$ being a representation matrix of $g \in C_{4v}$, where $s_i \tau_j$'s are the product of Pauli matrices acting on the spin, orbital, and sublattice degrees of freedom. Then, we obtain the total symmetric Hamiltonian \cite{kobayashi224504,yamazaki073701}.
(see Table \ref{IR} and Sec.~I of Supplemental Material \cite{Supp}):
\begin{align}
\nonumber
&\hat{H} = \frac{1}{2} \sum_{\vv k} \hat{\psi}^{\dagger}_{\vv k} H(\vv k) \hat{\psi}_{\vv k}, \\
&H(\vv k) = [v_1(\eta_1 k_y - \eta_2 k_x) + v_2(\eta_3 k_y-\eta_4 k_x)],
\label{toymodel}
\end{align}
where $\hat{\psi}^{\dagger}_{\vv k} = (\hat{\psi}^{\dagger}_{1\vv k},\hat{\psi}^{\dagger}_{2\vv k},\hat{\psi}^{\dagger}_{3\vv k},\hat{\psi}^{\dagger}_{4\vv k})$ and $\hat{\psi}_{\vv k} = {}^{t}(\hat{\psi}_{1\vv k},...,\hat{\psi}_{4\vv k})$ are the Majorana creation/annihilation operators, which satisfy $\hat{\psi}^{\dagger}_{1\vv k} = \hat{\psi}_{2-\vv k}, \hat{\psi}^{\dagger}_{3\vv k} = \hat{\psi}_{4-\vv k}$ and the subscripts $1,...,4$ denote four MFs.
$\eta_i$'s are given by $\eta_1 = (1/\sqrt{2})(s_1\tau_0+s_2\tau_0), \eta_2  = (-1/\sqrt{2})(s_1\tau_0 - s_2\tau_0), \eta_3 = (1/\sqrt{2})(s_1\tau_3+s_2\tau_3), \eta_4 = (-1/\sqrt{2})(s_1\tau_3-s_2\tau_3)$.
$H(\vv{k})$ satisfies $D_{g} H(\vv k) D^{\dagger}_{g} = H(g\vv k)$ where a momentum $\bm{k}$ is transformed to $g\bm{k}$ under the action of $g$.
The spin of MFs is represented by $\vv{\sigma} = (\sigma_x,\sigma_y,\sigma_z) \equiv (\eta_5,\eta_6,-s_3\tau_0)$ where $\eta_5 = (1/\sqrt{2})(s_1\tau_2 - s_2\tau_2), \eta_6 = (-1/\sqrt{2})(s_1\tau_2 + s_2\tau_2)$, which are coupled to the magnetization $\vb*{M}$ as $\vb*{M} \cdot \vb*{\sigma}$.
The time-reversal $\Theta$, particle-hole $C$, and chiral $\Gamma$ symmetries are defined by  $\Theta= s_2\tau_3 K, C= s_1\tau_0 K, \Gamma=s_3\tau_3$ where they satisfy $\Theta H(\vv k) \Theta^{-1} = H(-\vv k), \ C H(\vv k) C^{-1} = -H(-\vv k),\ \Gamma H(\vv k) \Gamma^{-1} = -H(\vv k)$.
In our model, which is the $C_{4v}$ PG symmetric model for the $A_{1u}$ bulk's superconducting pair potential with $O_h$ PG symmetry, the double MKPs have specific electromagnetic multipoles with particle-hole symmetry. Table \ref{IR} shows that the double MKPs have magnetic multipoles that are represented by $\bm{\sigma}$ and electric multipoles that are represented by $s_3 \tau_1$ with $B_1$ and $s_0\tau_2$ with $B_2$ representations. Therefore, as follows, we can define the spin currents $j^x_y +j^y_x$ and $j^x_x-j^y_y$ generated by dynamical strains $u_{xx}-u_{yy}$ and $u_{xy}+u_{yx}$, where $u_{ij}(\bm{x}) = \partial_i u_j(\bm{x})$ and $u_j(\bm{x}) \propto e^{i(\bm{q}\cdot \bm{x}-\omega t)}$ denote the displacement field, respectively (Fig. \ref{3D}).
For double MKPs, we can construct the spin current operator only from surface effective theory, i.e.,
the spin current is defined by $j^{\alpha}_{i} \equiv \frac{1}{2} \{ \sigma_{\alpha},\partial H(\vv{k})/\partial k_i \} \ (i=x,y)$, which satisfies particle-hole symmetry $C j^{\alpha}_{i} C^{-1} = -j^{\alpha}_{i}$.
Then, we obtain $\hat{j}^{x}_{y}(\vv{q}) + \hat{j}^{y}_{x}(\vv{q})= \frac{1}{2}\sum_{\vv{k}} \hat{\psi}^{\dagger}_{\vv{k}-\frac{\vv{q}}{2}}  [-v_2 s_3 \tau_1]  \hat{\psi}_{\vv{k}+\frac{\vv{q}}{2}}$ and $\hat{j}^{x}_{x}(\vv{q}) - \hat{j}^{y}_{y}(\vv{q})= \frac{1}{2}\sum_{\vv{k}} \hat{\psi}^{\dagger}_{\vv{k}-\frac{\vv{q}}{2}}  [v_1 s_0 \tau_2]  \hat{\psi}_{\vv{k}+\frac{\vv{q}}{2}}$.

%\textit{Majorana spin current generation by dynamical strain.---}
Here, we define the operators, which are represented by $\hat{O}^{(1)}(\vv{q}) \equiv \frac{1}{2} \sum_{\bm{k}} \hat{\psi}^{\dagger}_{\bm{k}-\frac{\bm{q}}{2}} (\rho_1 s_3\tau_1) \hat{\psi}_{\bm{k}+\frac{\bm{q}}{2}}$ and $\hat{O}^{(2)}(\vv{q}) \equiv \frac{1}{2} \sum_{\bm{k}} \hat{\psi}^{\dagger}_{\bm{k}-\frac{\bm{q}}{2}} (\rho_2 s_0\tau_2) \hat{\psi}_{\bm{k}+\frac{\bm{q}}{2}}$. Since $\hat{O}^{(1)}$ and $u_{xx}-u_{yy}$, $\hat{O}^{(2)}$ and $u_{xy}+u_{yx}$ share the same representation,
the couplings between double MKPs and the dynamical strains are represented by
\begin{align}
\nonumber
\hat{H}_{\text{surf,ex}} &= -\Bigg\{ \hat{O}^{(1)}(\vv{q}) [u_{xx}(\bm{q},\omega)-u_{yy}(\bm{q},\omega)]  \\ &+ \hat{O}^{(2)}(\vv{q}) [u_{xy}(\bm{q},\omega)+u_{yx}(\bm{q},\omega)] \Bigg\}. \label{strain}
\end{align}
Then, we calculate the linear response of the spin currents $\hat{j}^{x}_{y} + \hat{j}^{y}_{x}$ and $\hat{j}^{x}_{x} - \hat{j}^{y}_{y}$ to the dynamic strains given by Eq. (\ref{strain}):
\begin{align}
\nonumber
\langle \hat{j}^{x}_{y} + \hat{j}^{y}_{x} \rangle (\bm{q},\omega) &\equiv K_1(\vv{q},\omega)[u_{xx}(\bm{q},\omega)-u_{yy}(\bm{q},\omega)] \\
&\ \ \quad + K_2(\vv{q},\omega)[u_{xy}(\bm{q},\omega)+u_{yx}(\bm{q},\omega)], \\ \nonumber
\langle \hat{j}^{x}_{x} - \hat{j}^{y}_{y} \rangle (\bm{q},\omega) &\equiv K'_1(\vv{q},\omega)[u_{xx}(\bm{q},\omega)-u_{yy}(\bm{q},\omega)] \\
&\ \ \quad + K'_2(\vv{q},\omega)[u_{xy}(\bm{q},\omega)+u_{yx}(\bm{q},\omega)],
\end{align}
where the response function $K_{i}(\vv{q},\omega)$ is given by Ref. \cite{funato168436}
\begin{align}
K_{i}(\vv{q},\omega) \equiv i\int^{\infty}_{0} dt e^{i(\omega+i\delta)t} \ \langle [\hat{j}^{x}_{y}(\vv{q},t) + \hat{j}^{y}_{x}(\vv{q},t),\hat{O}^{(i)}(-\vv{q},0)] \rangle, \label{rf}
\end{align}
where $\hat{A}(t) = e^{i\hat{{H}}t} \hat{A} e^{-i \hat{H} t}$ and $\delta \to +0$. $K'_{i}(\vv{q},\omega)$ is also defined by replacing $\hat{j}^{x}_{y} + \hat{j}^{y}_{x}$ with $\hat{j}^{x}_{x} - \hat{j}^{y}_{y}$ in Eq. (\ref{rf}).
Note that the chemical potential $\mu$ for MFs is equal to $0$ due to particle-hole symmetry in superconducting states. Additionally, $\langle \cdots \rangle = \text{tr}[e^{-\hat H/T} \cdots]/\text{tr}[e^{-\hat H/T}]$.
If there are no applied external fields, $K_1(\vv{q},\omega)$ and $K'_2(\vv{q},\omega)$ then only take finite values since the spin currents and dynamic strains share the same representation of $C_{4v}$.
Here, we assume that the wavenumber $q$ and the frequency $\omega$ are smaller than the mean free path $l$ and relaxation time $\tau$ of the MFs.
These conditions are represented by $q \ll l^{-1}$, $\omega \ll \tau^{-1} =2\gamma$.
We expand the response function for $\omega$ and consider the nonequilibrium part: $K_i(\omega) - K_i(0)$.
(the details are shown in Sec.~II of the Supplemental Material \cite{Supp}):
\begin{align}
 K_1(\omega) - K_1(0)
\simeq \frac{i\omega}{8\pi^2} \frac{\rho_1}{v_1} \text{ln}&\left[\frac{(v_1-v_2)^2}{(v_1+v_2)^2}\right].
\label{results1}
\end{align}
One can see that $K_1(\omega) - K_1(0)$ is independent of the damping parameter $\gamma$.
Note that $K'_2(\omega) - K'_2(0) = \frac{v_1 \rho_2}{-v_2 \rho_1}(K_1(\omega) - K_1(0))$ because of chiral symmetry. The details are shown in Sec.~II of the Supplemental Material \cite{Supp}.

\begin{table*}
	\caption{Electromagnetic degrees of freedom of the double Majorana Kramers pairs (MKPs).
		They emerge on the surface with wallpaper-group (WG) symmetry, when the bulk pair potential belongs to irrep $\Delta$.
		Irreps of double MKPs, magnetic operators, electric operators of MKPs, which couple to magnetization $M_z$ perpendicular to the surface and strain [$u_{ij}(\boldsymbol{x}) = \partial_i u_j(\boldsymbol{x})$, where $\boldsymbol{u}(\boldsymbol{x})$ denotes the displacement field], are also shown. The last column denotes that the strain can couple to the double MKPs and the generated spin current by the strain.
		We adapt the definition for WG and irreps by Bilbao Crystallographic Server \cite{Elcoro}.
		$\bar\Gamma_i$ denotes the $i$th double-valued irrep of the little group on the $\Gamma$ point. $2\bar\Gamma_i = \bar\Gamma_i \oplus \bar\Gamma_i$.
        The result for $P3m1$ are the same as that for $P31m$.
	}
\begin{ruledtabular}
		\begin{tabular}{lllllccc}
			\multirow{2}{*}{WG} & \multirow{2}{*}{$\Delta$} & \multirow{2}{*}{MKPs} & \multirow{2}{*}{Magnetic} & \multirow{2}{*}{Electric} & \multirow{2}{*}{Magnetization $M_z$} & \multicolumn{2}{c}{Strain (Spin current)}
			\\
            & & & & & & $u_{xx}-u_{yy}$ ($j^{x}_y + j^{y}_x$) & $u_{xy}+u_{yx}$ ($j^{x}_x - j^{y}_y$)
            \\
			\hline
			$p4m$ & $A_2$ & $\bar\Gamma_6 \oplus \bar\Gamma_7$ & $2A_2+E$ & $B_1+B_2$ & gapped & $\bigcirc$ & $\bigcirc$
			\\
    		$p31m$ & $A_1$ &  $2(\bar\Gamma_4 \oplus \bar\Gamma_5)$ & $4A_1$ & $2A_2$ & gapless & \multicolumn{2}{c}{$\times, \text{ only coupled to } u_{xy}-u_{yx} \ (j^{x}_x + j^{y}_y)$}
            \\
            $p4g$ & $B_1$ &  $\bar M_6 \oplus \bar M_7$ & $A_2+B_1+E$ & $A_2+B_2$  & gapped & $\times, \text{ coupled to } u_{xy}-u_{yx} \ (j^{x}_x + j^{y}_y)$ & $\bigcirc$
 \end{tabular}	
\end{ruledtabular}
\label{result1}
\end{table*}

\begin{figure}[t]%\begin{figure*}
\hspace{1mm}
	\centering
\includegraphics[scale=0.3]{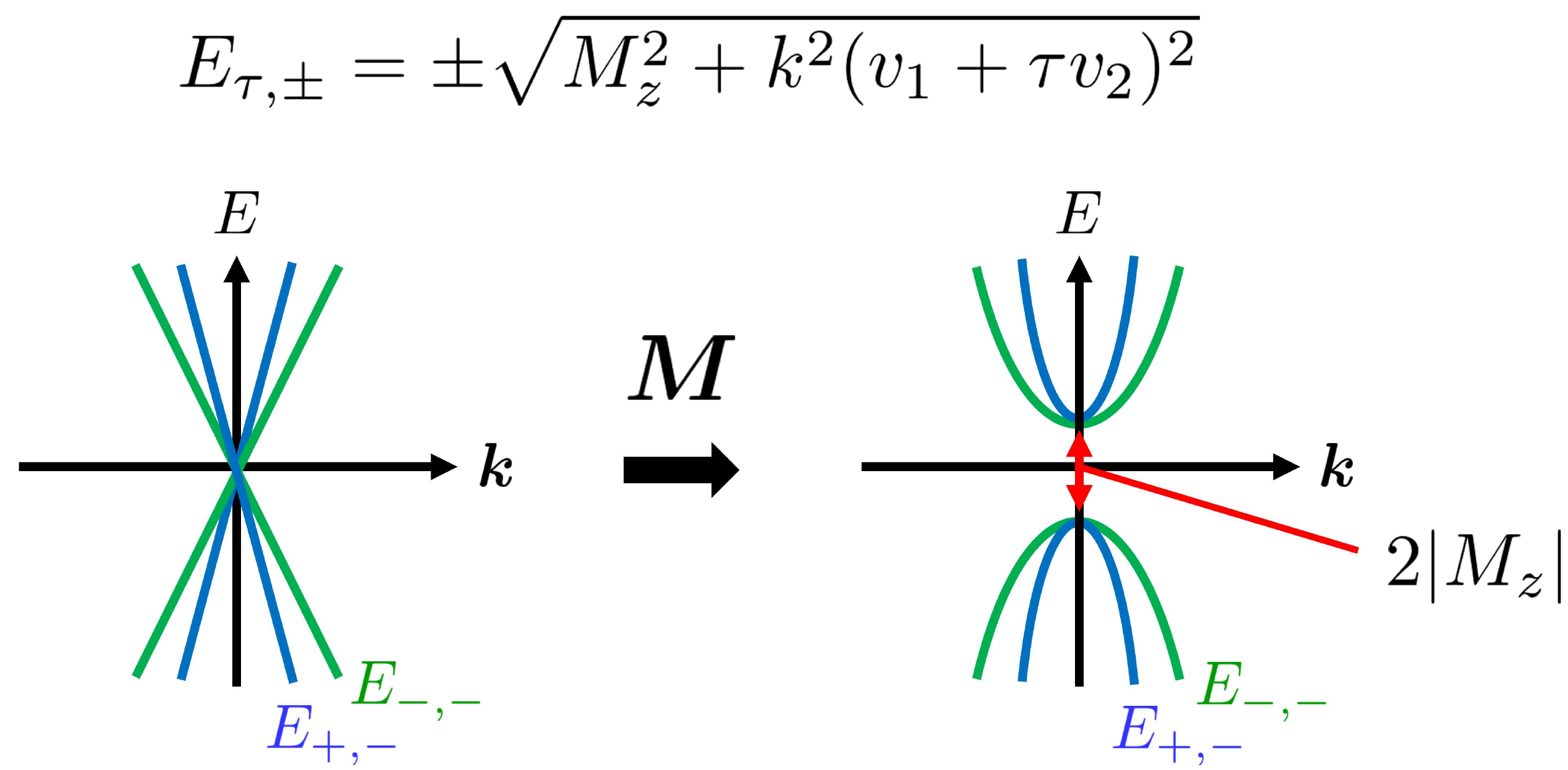}
\caption{The energy dispersion of (left) Eq.~(\ref{toymodel}) and (right) Eq.~(\ref{toymodel2}). The energy gap opened by the magnetization $\bm{M}$.}
\label{dispersion}
\vspace{-5.5mm}
\end{figure}%\end{figure*}

\begin{figure}%\begin{figure*}
\includegraphics[width=0.48\textwidth]{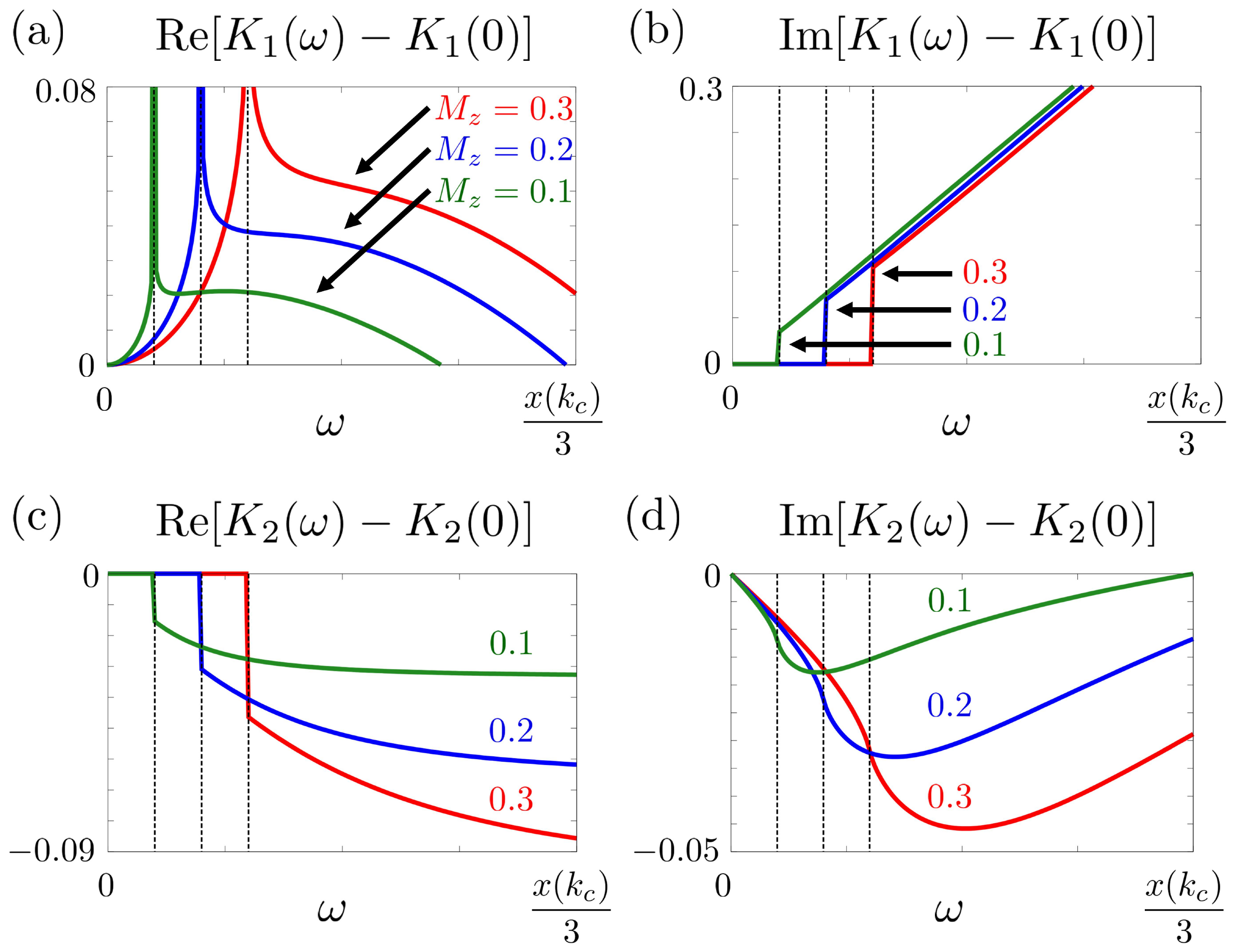}
\caption{The numerical results of Eqs. (\ref{LehmannK1}) and (\ref{LehmannK2}). (a) and (b) show the amplitudes of the real and imaginary parts of $K_1(\omega)-K_1(0)$, respectively. (c) and (d) show the real and imaginary parts of $K_2(\omega)-K_2(0)$, respectively. Red, blue and green lines denote the cases of $|M_z|=0.3$, $|M_z|=0.2$ and $|M_z|=0.1$, respectively. The values of the three black dotted lines drawn on the horizontal axis in (a), (b), (c), and (d) denote the size of the energy gap $2|M_z|$, and from left to right are $0.2$, $0.4$, and $0.6$. The parameters are $\rho_1 =1.0, \rho_2 =1.0, v_1 = 0.3, v_2 = 0.2, k_c =10$.}
\label{K}
\vspace{-5.5 mm}
\end{figure}

Next, we consider the case in which the dispersion of double MKPs is gapped.
This situation can be realized, for example, by attaching ferromagnets on the surface.
Then, the effective Hamiltonian is given by
\begin{align}
\tilde{H}(\vv k) = H(\vv k) + \tilde{M}_z \sigma_z.
\label{toymodel2}
\end{align}
The second term in Eq.~(\ref{toymodel2}) is the Zeeman term reflected by coupling between the spin moment of double MKPs and the magnetization of ferromagnets, where we define $M_z \equiv -\tilde{M}_z$. Then, the energy dispersion is given by $E_{\tau=\pm,\pm}=\pm \sqrt{M^2_z + k^2(v_1+\tau v_2)^2}$ in Fig.~\ref{dispersion}.
The applied magnetization opens an energy gap in the dispersion of double MKPs and lowers the symmetry from $C_{4v}$ to $C_4$. Therefore, $K_2(\omega)$ can take a finite value since ${j}^{x}_{y} + {j}^{y}_{x}$ and $u_{xy}+u_{yx}$ share the same irreducible representation of $C_{4}$.
The relationship between $B_z$ and $\omega$, is important; hence, we calculate the response function by using the Lehmann representation when the dispersion of double MKPs is gapped:
\begin{align}
\nonumber
K_1(\omega) &= \frac{1}{4}\sum_{n\in \text{occ}, m\in \text{unocc}} \Bigg\{\frac{\langle n|-v_2 s_3 \tau_1| m \rangle \langle m|\rho_1 s_3 \tau_1 |n \rangle}{E_n -E_m + \omega + i\delta} \\ &\ \quad+ \left[ \frac{\langle n|-v_2 s_3 \tau_1| m \rangle \langle m|\rho_1 s_3 \tau_1 |n \rangle}{E_n -E_m - \omega + i\delta} \right]^{*} \Bigg\}, \label{LehmannK1} \\
\nonumber
K_2(\omega) &= \frac{1}{4}\sum_{n\in \text{occ}, m\in \text{unocc}} \Bigg\{\frac{\langle n|-v_2 s_3 \tau_1| m \rangle \langle m|\rho_2 s_0 \tau_2 |n \rangle}{E_n -E_m + \omega + i\delta} \\ &\ \quad+ \left[ \frac{\langle n|-v_2 s_3 \tau_1| m \rangle \langle m|\rho_2 s_0 \tau_2 |n \rangle}{E_n -E_m - \omega + i\delta} \right]^{*} \Bigg\}.
\label{LehmannK2}
\end{align}
where $\delta \to +0$, $|n\rangle$ and $|m\rangle$ denote occupied states and unoccupied states of Eq. (\ref{toymodel2}), respectively.
As a result, we obtain the nonequilibrium part: $K_i(\omega) - K_i(0)$.
\begin{align}
\nonumber
&K_1(\omega)-K_1(0) \\
&=
 \begin{cases}
  A\frac{\omega^2}{M_z} & \hspace{-35mm}(\omega \ll 2|M_z|),
  \\
  B\left(- \frac{\omega}{2}\text{ln}\left| \frac{1+\frac{2k_c v_1}{\omega}}{1-\frac{2k_c v_1}{\omega}} \right| + \frac{2M_z v^2_1}{v^2_1-v^2_2} - \frac{M_z v_1 \text{ln}\big|\frac{v_1-v_2}{v_1+v_2}\big|}{2v_2}\right) + i\frac{B \pi v_1}{2v_2}\omega \\
& \hspace{-35mm} (2|M_z| \ll \omega \ll x(k_c)),
 \end{cases}
\label{ResultsK1}
\\ \nonumber
&K_2(\omega) - K_2(0) \\
&=
 \begin{cases}
  i C \omega & (\omega \ll 2|M_z|),
  \\
  D \pi M_z +i D M_z\text{ln}\Bigg|\frac{1+\frac{2 k_c v_1}{\omega}}{1-\frac{2 k_c v_1}{\omega}} \Bigg| & (2|M_z| \ll \omega \ll x(k_c)),
 \end{cases}
\label{ResultsK2}
\\ \nonumber &A=\frac{2v_1v_2(3v^2_2-v^2_1)-((v^2_1+v^2_2)^2-4v^4_2)\text{ln}\big|\frac{v_1-v_2}{v_1+v_2}\big|}{32 v^3_1 v^3_2}, \ B=\frac{v_2\rho_1}{8\pi v^2_1},
\\ &C=\frac{v_2\rho_2}{4\pi} \frac{-2v_1 v_2 + v^2 \text{ln}\Big|\frac{v_1+v_2}{v_1-v_2}\Big|}{8v^2_1 v^2_2}, \ D=-\frac{v_2\rho_2}{8\pi}\frac{v_2}{v^3_1-v_1 v^2_2},
\end{align}

\noindent
where $x(k_c)=\sqrt{M^2_z+k^2_c(v_1+v_2)^2}+\sqrt{M^2_z+k^2_c(v_1-v_2)^2}$ and $k_c$ is a cutoff value (see Sec.~II of the Supplemental Material for detailed results \cite{Supp}).
Eqs. (\ref{ResultsK1}) and (\ref{ResultsK2}) show that $K_1(\omega)$ and $K_2(\omega)$ have two frequency regions with different behaviors.
The numerical results of Eqs.~(\ref{LehmannK1}) and (\ref{LehmannK2}) are shown in Fig.~\ref{K}, and they reproduce the approximate analytical solutions obtained in each region well.
Both $K_1(\omega)-K_1(0)$ and $K_2(\omega)-K_2(0)$ reflect the energy dispersion of the MFs. In particular, there are abrupt changes in the values of (a), (b), and (c) in Fig.~\ref{K}, which are the points at $\omega=2|M_z|$, just when the frequency is equal to the band gap.
On the other hand, (d) is the diagram corresponding to the result of the principal value integration in Eq. (\ref{LehmannK2}), and there is no peak structure even at $\omega=2|M_z|$ where the density of states becomes finite. This is because the chemical potential of MFs is exactly zero, and $K_2(\omega)-K_2(0)$ is the quantity that can take a finite value when the dispersion of MFs exhibits an energy gap with applied magnetization. Hence, the contribution from the Fermi sea is considered to be dominant.
%\textcolor{red}{At $\omega \ll 2|M_z|$, $K_1(\omega)$ is proportional to $\text{ln}(k_c)$ since the impurities are neglected. In fact, by the same calculation that derives Eq. (\ref{results  1}), we find that $K^{(1)}_1 \propto \gamma^2/(M^2_z + \gamma^2)$. [???]}

%\end{figure*}

%\textit{Discussion.---}
In this letter, we show that the Majorana spin current generated by spatially nonuniform dynamic strain flows on the (001) surface of the 3D time-reversal-invariance TCSC Sr$_3$SnO.
For the gapless case, the spin current does not show damping dependence due to the linear dispersion of the double MKPs and the fact that their chemical potential is exactly zero.
In contrast, for the gapped case, the spin currents reflect the energy dispersion of the MFs, and they have a unique frequency dependence.
We emphasize again that the electromagnetic multipoles formed by double MKPs determine the properties of the spin current responses to dynamic strains.
%In our model, these multipoles belong to $B_1$ and $B_2$ representations since we consider the $C_{4v}$ PG symmetric model for the $A_{1u}$ bulk's superconducting pair potential with $O_h$ PG symmetry.
To see this, we show the results of the previous study \cite{yamazaki073701} in Table \ref{result1} (See the Supplemental material of Ref. \cite{yamazaki073701} for the complete version). This table shows the electromagnetic degrees of freedom of double MKPs for several WGs, irreps of the pair potential, and double MKPs. The case of $p4m$ corresponds to the present model. In the case of $p4m$ and $p4g$, the applied magnetization $M_z$ induces the energy gap of double MKPs; however, in the case of $p31m$, the double MKPs still remain gapless. In addition, $p4m$ and $p4g$ are the same gapped case, but in the case of $p4g$, the double MKPs do not couple to the strain $u_{xx}-u_{yy}$ but couple to $u_{xy}-u_{yx}$ owing to the protection of glide-plane symmetry. We have specifically shown the application of the general effective theory for the electromagnetic properties of surface double MKPs derived from system symmetries, such as crystalline and superconducting symmetries, and the new dynamics of MFs characterized by these symmetries. We believe that our work accelerates the study of the coupling of given external fields with MFs and the transport phenomena arising therefrom.

As a detection method for AC spin currents, spin wave resonance can be observed by injecting spin currents into ferromagnetic metals \cite{Kobayashi077202,Tateno104406} and the rectifying effect of magnetostriction \cite{Kawada9697}. The dynamic strains $u_{xx}(\bm{r},t)$ and $u_{xy}(\bm{r},t)$ can be realized by a Rayleigh wave and a Love wave propagating in the $x$-direction, respectively.
In relation to the experiment, parameters such as $\rho_1,\rho_2,v_1,v_2$ can be obtained by projecting bulk operators onto the surface. This study will be published in a future paper.

The authors gratefully acknowledge M. Matsuo, Y. Nozaki, J. J. Nakane, Y. Imai, T. Yamaguchi, T. Matsushita, R. Kikuchi, and H. Kohno for valuable discussions.
This work was supported by JSPS KAKENHI for Grants (No. JP20K03835) and the Sumitomo Foundation (190228).
Y.Y. is supported by Grant-in-Aid for JSPS Fellows Grant No. 22J14452.
This work was financially supported by JST SPRING, Grant Number JPMJSP2125. The author Y.Y. would like to take this opportunity to thank the ``Interdisciplinary Frontier Next-Generation Researcher Program of the Tokai Higher Education and Research System.''

%\bibliography{ref}

%Formatter's Note: Your references and in-text citations have been formatted to conform to the journal's guidelines. We encourage you to keep these changes.

%

\end{document}